# On Zero-Error Source Coding with Feedback

Mayank Bakshi    Michelle Effros
Department of Electrical Engineering
California Institute of Technology
Pasadena, California 91125, USA
Email: {mayank, effros}@caltech.edu

*Abstract*—We consider the problem of zero error source coding with limited feedback when side information is present at the receiver. First, we derive an achievable rate region for arbitrary joint distributions on the source and the side information. When all source pairs of source and side information symbols are observable with non-zero probability, we show that this characterization gives the entire rate region. Next, we demonstrate a class of sources for which asymptotically zero feedback suffices to achieve zero-error coding at the rate promised by the Slepian-Wolf bound for asymptotically lossless coding. Finally, we illustrate these results with the aid of three simple examples.

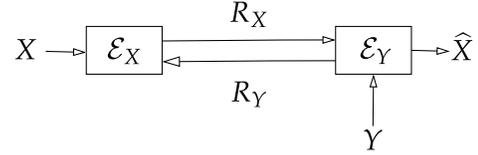

Fig. 1. Source Coding with feedback

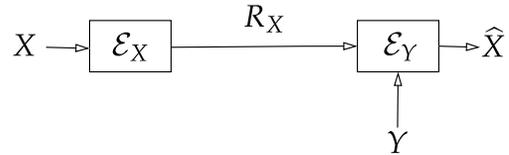

Fig. 2. Source Coding without feedback

## I. INTRODUCTION

Traditionally, in most systems that are considered in Network Source Coding literature, the flow of information is assumed to be unidirectional. However, networks that occur in practice often comprise of bidirectional links. For example, in a network with wireless links, it is possible for any pair of nodes to act as a transmitter-receiver pair. In such networks, if there is only one source and sink, then assuming a unidirectional flow of information is not restrictive from a source coding point of view as transmission from the sink to the source cannot supply the source with any useful information. In contrast, if the network consists of multiple sources or sinks, then, feedback from sinks to sources has the potential to alter the forward rate region [1].

Unfortunately, fully characterizing the rate tradeoffs for networks with bidirectional links is a non-trivial problem. In part, the difficulty arises from the fact that in such networks, potentially unbounded number of transmissions may be required to achieve optimal rates [2], and usual information theoretic techniques do not readily extend to these situations.

In this paper, we aim to develop insights into these systems by studying a simple network. The setup that we consider is shown in Figure 1. The source process $X$ is observed at $\mathcal{E}_x$ and demanded at $\mathcal{E}_y$. Terminal $\mathcal{E}_y$ observes source $Y$ jointly distributed with $X$. We wish to characterize the set of rates $(R_X, R_Y)$ required to enable $\mathcal{E}_y$ to reconstruct $X$ with precisely zero probability of error.

A relaxed version of this problem is the asymptotically lossless setting, where the process $X$ is demanded with a vanishing error probability as the block length increases without bound. For this setting, it is easy to see that even unlimited rate on the feedback link does not change the rate required. In particular, for the system shown in Figure 2, Slepian and Wolf [3] have shown that the minimum rate required on the forward link equals the cutset bound. Since the addition of the feedback link does not alter the cutset bound, it follows that the rate required cannot be reduced any further.

In the zero-error setting, the cutset bound is not achievable for the system shown in Figure 2 (c.f. [4],[5],[6],[7]) since $\mathcal{E}_x$ has to distinguish between all possible pairs of source and side information sequences, not just those that are typical. We show that a two-way communication between $\mathcal{E}_x$ and $\mathcal{E}_y$ allows them to decide whether or not their observed sequences are typical, and enables a tradeoff between the rate on the forward and the backward links. Of special interest are two extreme cases that are discussed in Examples 1 and 2. In the first example, asymptotically zero rate on the feedback link enables $\mathcal{E}_x$ to operate at rates arbitrarily close to the cutset bound, while in the second, the sum rate required on the two links is bounded from below by $H(X)$.

It should be noted that the study of feedback under the zero error criterion is not entirely new. Prior works on communication complexity have examined some aspects of this problem (see for example, [8], [9], [10], [11]). In this work, we combine insights from both communication complexity theory and asymptotically lossless source coding to show that feedback is useful in the zero error setting, even for networks where it does not help in the asymptotically zero error setting.

The rest of the paper is organized as follows. The main results are presented in Section III. Theorem 1 gives an achievable rate region for the setup in Fig 1. Theorems 2 and 3 give the exact rate region when the joint distribution of $X$ and $Y$ satisfy certain conditions. Finally, in Section IV, we examine a few examples to illustrate the main results.

[0]This material is based upon work partially supported by DARPA IT-MANET and Caltech's Lee Center for Advanced Networking.

We begin with defining the notation in Section II.

## II. PRELIMINARIES

Let $p_{XY}$ be a probability mass function on a finite alphabet $\mathcal{X} \times \mathcal{Y}$. Denote by $p_X$ (resp. $p_Y$), the marginal of $p_{XY}$ on $\mathcal{X}$ (resp. $\mathcal{Y}$). For each $n \in \mathbb{N}$, the collection of random variable $(X_1, Y_1), (X_2, Y_2), \ldots, (X_n, Y_n)$ is drawn i.i.d. from the distribution $p_{XY}$. Let $\{0, 1\}^*$ denote the set of all finite length sequences drawn from $\{0, 1\}$. Let $l : \{0, 1\}^* \to \mathbb{N}$ be the length function on $\{0, 1\}^*$, i.e. if $b$ is a string of $n$ bits, then $l(b) = n$.

Define an $n$-dimensional, $k$-interactive code $(f, g)$ to be a collection of $2k - 1$ functions

$$f_i : \mathcal{X}^n \times \mathcal{Y}^n \to \{0, 1\}^* \text{ for } i = 1, 2, \ldots, k,$$
and $$g_i : \mathcal{X}^n \times \mathcal{Y}^n \to \{0, 1\}^* \text{ for } i = 1, 2, \ldots, k - 1$$

that satisfy the property that for each $(\mathbf{x}, \mathbf{y}) \in \mathcal{X}^n \times \mathcal{Y}^n$, $f_1(\mathbf{x}, \mathbf{y}) = \widetilde{f}_1(\mathbf{x})$, and for $i = 1, 2, \ldots, n - 1$, $f_{i+1}(\mathbf{x}, \mathbf{y}) = \widetilde{f}_{i+1}(\mathbf{x}, g_1(\mathbf{x}, \mathbf{y}), \ldots, g_i(\mathbf{x}, \mathbf{y}))$ and $g_i(\mathbf{x}, \mathbf{y}) = \widetilde{g}_i(\mathbf{y}, f_1(\mathbf{x}, \mathbf{y}), \ldots, f_i(\mathbf{x}, \mathbf{y}))$ for some collection of functions $\{(\widetilde{f}_i, \widetilde{g}_i) : i = 1, \ldots, n - 1\}$. We call blocklength-$n$, $k$-interactive code $(f, g)$ a zero-error code if there exists a decoder function $h : (\{0, 1\}^*)^k \times \mathcal{Y}^n \to \mathcal{X}^n$ such that for all $(\mathbf{x}, \mathbf{y}) \in \mathcal{X}^n \times \mathcal{Y}^n$, $h(f_1(\mathbf{x}, \mathbf{y}), f_2(\mathbf{x}, \mathbf{y}), \ldots, f_k(\mathbf{x}, \mathbf{y}), \mathbf{y}) = \mathbf{x}$. We allow $(f, g)$ to be a variable-length code. The average rate for the code thus defined is the pair $(R_X, R_Y)$, where

$$R_X = \frac{1}{n} \sum_{i=1}^{k} \mathbf{E}[l(f_i(X^n, Y^n))]$$
and $$R_Y = \frac{1}{n} \sum_{i=1}^{k-1} \mathbf{E}[l(g_i(X^n, Y^n))].$$

We say that a rate point $(R_X, R_Y)$ is zero-error achievable if, for some integers $n$ and $k$, there exists an block length-$n$, $k$-interactive zero-error code with average rate $(R_X, R_Y)$. The rate region $\mathcal{R}_Z(X, Y)$ is the closure of the set of all zero-error achievable rates. Let $A_\epsilon^{(n)}(X)$ denote the set of $\epsilon$-strongly typical sequences in $\mathcal{X}^n$. Similarly, define $A_\epsilon^{(n)}(Y)$ and $A_\epsilon^{(n)}(X, Y)$. The type class of probability mass function $Q$ is denoted by $T^{(n)}(Q)$ (see [12] for details).

## III. RESULTS

We derive an achievable region in Theorem 1. Towards this end, we first present a weaker version of the theorem in the following Lemma.

*Lemma 1:* $\mathcal{R}_Z(X, Y) \supseteq \{(R_X, R_Y) : R_X \geq H(X|Y), R_X + R_Y \geq H(X)\}$

*Proof:* Let $R > H(X|Y)$. Consider the following code construction.

Fix a block length $n \geq 1$. Partition $A_\epsilon^{(n)}(X)$ into $2^{nR}$ bins $\{\mathcal{B}_i : i = 1, 2, \ldots, 2^{nR}\}$ by assigning each $\mathbf{x} \in A_\epsilon^{(n)}(X)$ a bin chosen uniformly at random. Let $B : A_\epsilon^{(n)}(X) \to \{1, 2, \ldots, 2^{nR}\}$ denote the mapping from sequences in $\mathcal{X}^n$ to the corresponding bin number. For $i = 1, 2, \ldots, 2^{nR}$, let $I^i : \mathcal{B}_i \to \{1, 2, \ldots, |\mathcal{B}_i|\}$ be a numbering of sequences in the $i$-th bin.

Let $\mathbf{x}$ and $\mathbf{y}$ be the sequences observed by $\mathcal{E}_x$ and $\mathcal{E}_y$ respectively. Consider the block length-$n$, 2-interactive code that defines the following protocol:

1. $\mathcal{E}_x$ sends $f_1(\mathbf{x}, \mathbf{y})$, where

$$f_1(\mathbf{x}, \mathbf{y}) = \begin{cases} 0 \cdot \mathbf{x} & \text{if } \mathbf{x} \notin A_\epsilon^{(n)}(X) \\ 1 \cdot B(\mathbf{x}) & \text{otherwise.} \end{cases}$$

2. If $f_1(\mathbf{x}, \mathbf{y}) = 0 \cdot \mathbf{x}$, then the procedure stops. Else,
   a. $\mathcal{E}_y$ sends

$$g_1(\mathbf{x}, \mathbf{y}) = \begin{cases} 0 & \text{if } \mathbf{y} \notin A_\epsilon^{(n)}(Y) \text{ or} \\ & (\widetilde{\mathbf{x}}, \mathbf{y}) \notin A_\epsilon^{(n)}(X, Y) \\ & \forall \, \widetilde{\mathbf{x}} \in \mathcal{B}_{B(\mathbf{x})} \\ 1 \cdot I^{B(\widehat{\mathbf{x}})} & \text{otherwise} \end{cases}$$

where,
$$\widehat{x} = \arg \min_{\widehat{\mathbf{x}} \in \mathcal{B}_{B(\mathbf{x})}} I_{B(\mathbf{x})}(\widehat{\mathbf{x}}).$$

   b. $\mathcal{E}_x$ sends

$$f_2(\mathbf{x}, \mathbf{y}) = \begin{cases} 0 \cdot \mathbf{x} & \text{if } g_1(\mathbf{x}, \mathbf{y}) = 0 \text{ or} \\ & g_1(\mathbf{x}, \mathbf{y}) \neq 0 \text{ and} \\ & I^{B(\mathbf{x}(\widehat{\mathbf{x}})} \neq I^{B(\mathbf{x}(\mathbf{x})} \\ 1 & \text{otherwise.} \end{cases}$$

Since the mapping from $\mathbf{x}$ to the pair $(B(\mathbf{x}), I^{B(\mathbf{x})}(\mathbf{x}))$ is one-to-one, the above protocol ends with $\mathcal{E}_y$ decoding $\mathbf{x}$ correctly for each $\mathbf{x} \in \mathcal{X}^n$. Let $P_n$ denote the probability that the sequence of transmissions is

$$1 \cdot B(\mathbf{x}); \ 1 \cdot I^b(\mathbf{x}); \ 1$$

Let $R_X^{(n)}$ and $R_Y^{(n)}$ denote the expected rates on the forward link and the backward link respectively. These can be bounded from above as

$$R_X^{(n)} \leq P_n(R + 2/n) + (1 - P_n)(H(X) + 2/n)$$
and $$R_Y^{(n)} \leq P_n((1/n)\mathbf{E}[\log |\mathcal{B}_{B(X^n)}|]) + (1 - P_n)(1/n)$$
$$= P_n((H(X) - R + \epsilon) + (1 - P_n)(1/n).$$

Following previous results on random binning (c.f.[3]), it is easily seen that $P_n$ approaches 1 as $n$ grows without bound. Thus,

$$\limsup_{n \to \infty} R_X^{(n)} \leq R$$
and $$\limsup_{n \to \infty} R_Y^{(n)} \leq H(X) - R + \epsilon.$$

Since $\epsilon$ is arbitrary, this proves the desired result. ∎

Next, using previous results on zero-error coding without feedback [4] (See Fig 2), we improve the above rate region. Let $H_Z(X|Y)$ denote the minimum rate required for describing $X$ without error when $Y$ is known at the decoder. The following theorem gives an achievable region when a feedback link is present.

*Theorem 1:* $\mathcal{R}_Z(X,Y) \supseteq \{(R_X, R_Y) : R_X \geq H(X|Y), R_X + R_Y \geq H_Z(X|Y)\}$.

*Proof:* Let $R > H_Z(X|Y)$. By the result of [4], for some block length $n$, there exists a function $c : \mathcal{X}^n \to \mathcal{C}$ satisfying the following properties:

1) Let $\mathbf{x}, \mathbf{x}' \in \mathcal{X}^n$ such that there exists $\mathbf{y} \in \mathcal{Y}^n$ for which $p_{XY}(\mathbf{x}, \mathbf{y}) > 0$ and $p_{XY}(\mathbf{x}', \mathbf{y}) > 0$. Then, $c(\mathbf{x}) \neq c(\mathbf{x}')$.
2) $H(c(X^n)) = nR$.

Observe that knowing $c(\mathbf{x})$ is sufficient for $\mathcal{E}_y$ to decode $\mathbf{x}$ with zero error. Therefore, $\mathcal{R}_Z(X,Y) \supseteq \frac{1}{n}\mathcal{R}_Z(c(X^n), Y^n)$. Further, $H(c(X^n)|Y^n)) > nH(X|Y)$ from Slepian and Wolf's result [3].

Using Lemma 1, $\mathcal{R}_Z(c(X^n), Y^n) \supseteq \{(R_X, R_Y) : R_X \geq H(c(X^n)|Y^n), R_X + R_Y \geq H(c(X^n))\}$. It follows that $\mathcal{R}_Z(X,Y) \supseteq \{(R_X, R_Y) : R_X \geq H(X|Y), R_X + R_Y \geq H(X)\}$. ∎

Theorem 2 shows that Theorem 1 is tight for all $P_{XY}$ such that $p_{XY} > 0$ for all $(x,y) \in \mathcal{X} \times \mathcal{Y}$.

*Theorem 2:* Let $X \in \mathcal{X}$ and $Y \in \mathcal{Y}$ be random variables such that $p_{XY}(x,y) > 0$ for all $(x,y) \in \mathcal{X} \times \mathcal{Y}$. Then, $\mathcal{R}_Z(X,Y) = \{(R_X, R_Y) : R_X \geq H(X|Y), R_X + R_Y \geq H(X)\}$

*Proof:* The achievability of the given rates follow from Theorem 1. For the converse, we use an argument inspired by communication complexity theory (c.f. [9]). Let $(f,g)$ be an $n$-dimensional interactive code over $k$-sessions operating at the rate $(R_X, R_Y)$. Define the set $\mathcal{C}(f,g)$ to be the set of all possible codewords, i.e.,

$$\mathcal{C}(f,g) \triangleq \{(f,g)(\mathbf{x},\mathbf{y}) : (x,y) \in \mathcal{X}^n \times \mathcal{Y}^n\},$$

and let $\mathcal{D}(f,g)$ be a partition of $\mathcal{X}^n$ into sets that are inverse images of singletons in $\mathcal{C}(f,g)$, i.e.,

$$\mathcal{D}(f,g) = \{\{(\mathbf{x},\mathbf{y}) \in \mathcal{X}^n \times \mathcal{Y}^n : (f,g)(\mathbf{x},\mathbf{y}) = c\} : c \in \mathcal{C}(f,g)\}$$

Notice that our definition of an interactive code implies that if for some $D \in \mathcal{D}(f,g)$, $(\mathbf{x}_1, \mathbf{y}_1) \in D$ and $(\mathbf{x}_2, \mathbf{y}_2) \in D$, then $(\mathbf{x}_1, \mathbf{y}_2) \in D$ and $(\mathbf{x}_2, \mathbf{y}_1) \in D$ as well. Further, at the end of the transmission, $X^n$ has be decoded at $\mathcal{E}_y$ without error. Thus, whenever $(\mathbf{x}_1, \mathbf{y}) \in D$ and $(\mathbf{x}_2, \mathbf{y}) \in D$ for some $D \in \mathcal{D}(f,g)$, then $\mathbf{x}_1 = \mathbf{x}_2$. Therefore, every $D \in \mathcal{D}(f,g)$ is of the form $\{\mathbf{x}\} \times D_\mathbf{x}$ for some $\mathbf{x} \in \mathcal{X}^n$ and $D_\mathbf{x} \subseteq \mathcal{Y}^n$. For $\mathbf{x} \in \mathcal{X}^n$, let

$$\mathcal{D}_\mathbf{x}(f,g) \triangleq \{D \in \mathcal{D}(f,g) : D = \{\mathbf{x}\} \times D_\mathbf{x} \text{ for some } D_\mathbf{x} \subseteq \mathcal{Y}^n\}.$$

For the code $(f,g)$, let $R_X^{(n)}$ and $R_Y^{(n)}$ denote the expected rates on the forward link and the backward link respectively. Since $\mathcal{C}(f,g)$ is a uniquely decodable code over the input alphabet $\mathcal{D}(f,g)$, from the converse to the source coding theorem it

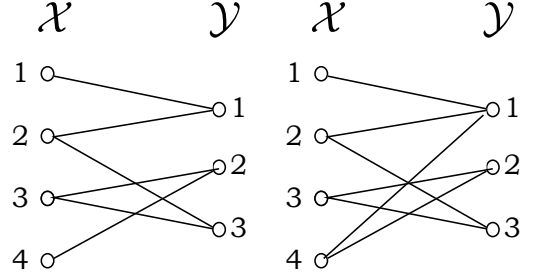

(a) A cycle-free p.m.f.   (b) A a p.m.f. with a cycle

Fig. 3. Example of $G(P_{XY})$ for two different distributions

follows that

$$\begin{aligned}
R_X^{(n)} + R_Y^{(n)} &\geq \frac{1}{n} \sum_{D \in \mathcal{D}(f,g)} p_{X^n Y^n}(D) \log \frac{1}{p_{X^n Y^n}(D)} \\
&= \frac{1}{n} \sum_{\mathbf{x} \in \mathcal{X}^n} \sum_{D \in \mathcal{D}_\mathbf{x}(f,g)} p_{X^n Y^n}(D) \log \frac{1}{p_{X^n Y^n}(D)} \\
&\geq \frac{1}{n} \sum_{\mathbf{x} \in \mathcal{X}^n} p_{X^n Y^n}(\{\mathbf{x}\} \times \mathcal{Y}) \log \frac{1}{p_{X^n Y^n}(\{\mathbf{x}\} \times \mathcal{Y})} \\
&= H(X)
\end{aligned}$$

by the concavity of entropy. Finally, note that even if $Y$ is known at $\mathcal{E}_x$, for successful decoding, we require the forward rate to be at least as large as $H(X|Y)$. Thus, $R_f \geq H(X|Y)$. ∎

For a probability mass function $P_{XY}$ on $\mathcal{X} \times \mathcal{Y}$, define $G(P_{XY})$, the connectivity graph of $X$ and $Y$, as a graph with vertices $\mathcal{X} \cup \mathcal{Y}$ and edges $\{(x,y) : p_{XY}(x,y) > 0\}$. We say that $P_{XY}$ is cycle-free if $G(P_{XY})$ has no cycles. See Fig 3 for an example of such a probability mass function. We next show that the rate $(R_X, R_Y) = (H(X|Y), 0)$ is in the zero error region $\mathcal{R}_Z(X,Y)$ when $P_{XY}$ is cycle-free.

*Theorem 3:* Let $X$ and $Y$ be random variables drawn from a joint distribution $P_{XY}$ that is cycle-free. Then, $\mathcal{R}_Z(X,Y) = \{(R_X, R_Y) : R_X \geq H(X|Y), R_Y \geq 0\}$.

*Proof:* The converse follows immediately from the Slepian-Wolf problem since the rate required on the forward link for zero-error coding is no less than the rate required for the asymptotically lossless Slepian-Wolf code. We now show the achievability of the claimed rates.

As in the proof of Theorem 1, let $R > H(X|Y)$ and partition $A_\epsilon^{(n)}(X)$ into $2^{nR}$ bins $\{\mathcal{B}_i : i = 1, 2, \ldots, 2^{nR}\}$ by assigning each $\mathbf{x} \in A_\epsilon^{(n)}(X)$ a bin chosen uniformly at random. Let $B : A_\epsilon^{(n)}(X) \to \{1,2,\ldots,2^{nR}\}$ denote the corresponding mapping from sequences in $A_\epsilon^{(n)}(X)$ to bin numbers. Let $\mathbf{x}$ and $\mathbf{y}$ be $n$-length sequences observed at $\mathcal{E}_x$ and $\mathcal{E}_y$ respectively. Denote the empirically observed type class of $\mathbf{x}$ by $\widehat{T}^{(n)}(\mathbf{x})$. Note that knowing $\widehat{T}^{(n)}(\mathbf{x})$ is sufficient for $\mathcal{E}_y$ to determine if $(\mathbf{x}, \mathbf{y}) \in A_\epsilon^{(n)}(X,Y)$.

Consider the following protocol.

1. $\mathcal{E}_x$ sends $f_1(\mathbf{x}, \mathbf{y}) = \widehat{T}^{(n)}(\mathbf{x})$.

2. $\mathcal{E}_y$ sends

$$g_1(\mathbf{x}, \mathbf{y}) = \begin{cases} 1 & \text{if } (\mathbf{x}, \mathbf{y}) \in A_\epsilon^{(n)}(X, Y) \\ 0 & \text{otherwise.} \end{cases}$$

3. $\mathcal{E}_x$ sends

$$f_2(\mathbf{x}, \mathbf{y}) = \begin{cases} B(\mathbf{x}) & \text{if } g_1(\mathbf{x}, \mathbf{y}) = 1 \\ \mathbf{x} & \text{otherwise} \end{cases}$$

4. If there is a unique $\mathbf{x}' \in \mathcal{B}_{B(\mathbf{x})}$ such that $(\mathbf{x}', \mathbf{y}) \in A_\epsilon^{(n)}(X, Y)$, or if $g_1(\mathbf{x}, \mathbf{y}) = 0$, transmission stops. Otherwise, $\mathcal{E}_y$ sends $g_2(\mathbf{x}, \mathbf{y}) = 0$.

5. $\mathcal{E}_x$ sends $f_3(\mathbf{x}, \mathbf{y}) = \mathbf{x}$

From Lemma 2, it follows that given individual type classes of $\mathbf{x}$ and $\mathbf{y}$, the joint type class is uniquely determined. Therefore, the above protocol always outputs the correct value $\mathbf{x}$. Finally using the same argument as in Theorem 1, as long as $R > H(X|Y)$, the expected rate on the forward link for the above code approaches $R$ as $n$ grows without bound, while the rate on the backward link approaches 0. ∎

## IV. DISCUSSION

We have shown that for every pair $(X, Y)$ such that the rate $H(X|Y)$ on the forward link is not achievable without feedback, the addition of the feedback link enables us to lower the forward transmission rate. In particular, for certain classes of sources, Theorem 3 shows that even asymptotically zero feedback is useful. The following example illustrates this.

*Example 1 (Binary Erasure Channel):* Let $X$ be distributed uniformly on $\mathcal{X} = \{0, 1\}$ and let $Y$ distributed on $\mathcal{Y} = \{0, E, 1\}$ with the transition probability

$$p_{Y|X}(y|x) = \begin{cases} 1 - p & \text{if } y = x \\ p & \text{if } y = e. \end{cases}$$

From prior results (c.f. [4]), it follows that without feedback, the minimum rate for zero error coding of $X$ is $H(X) = 1$. On the other hand, Theorem 3 shows that even with asymptotically zero feedback, a rate of $H(X|Y) = p$ is achievable on the forward link.

An interesting contrast to the above example is provided by the following example.

*Example 2 (Binary Symmetric Channel):* Let $X$ be distributed uniformly on $\mathcal{X} = \{0, 1\}$ and let $Y$ be distributed on $\mathcal{Y} = \{0, 1\}$ with the following transition probability

$$p_{Y|X}(y|x) = \begin{cases} 1 - p & \text{if } y = x \\ p & \text{if } y \neq x. \end{cases}$$

The minimum rate possible without feedback for this example is the same as that in Example 1. However, the presence of asymptotically zero feedback does not reduce the minimum rate required on the forward link. However, Theorem 2 enables using non-zero rate on the feedback link to operate at lower rates on the forward link. In particular, $\mathcal{R}_z(X, Y)$ is given by $\{(R_X, R_Y) : R_X \geq H(p), R_X + R_Y \geq 1\}$.

Finally, note that it is not $P_{XY}$ being cycle free is not a necessary condition for Theorem 3 to hold. This is shown in the following example.

*Example 3 (Binary Erasure Channel with Two Erasures):* Let $X$ be distributed uniformly on $\mathcal{X} = \{0, 1\}$ and let $Y$ distributed on $\mathcal{Y} = \{0, E_1, E_2, 1\}$ with the transition probability

$$p_{Y|X}(y|x) = \begin{cases} 1 - p & \text{if } y = x \\ p/2 & \text{if } y = e_1 \text{ or } e_2. \end{cases}$$

Lemma 3, which is proved in the Appendix shows that this example, is in fact, equivalent to Example 1. Thus, a rate $H(X|Y) = p$ on the forward link can be achieved with asymptotically zero rate on the feedback link, even though $P_{XY}$ is not cycle free.

## APPENDIX

*Lemma 2:* Let $T^{(n)}(Q_X) \subseteq \mathcal{X}^n$, $T^{(n)}(Q_Y) \subseteq \mathcal{Y}^n$, and $T^{(n)}(Q_{XY}) \subseteq \mathcal{X}^n \times \mathcal{Y}^n$ be type classes that are consistent with each other, i.e., the marginal of $Q_{XY}$ on $\mathcal{X}$ (resp. $\mathcal{Y}$) is $Q_X$ (resp. $Q_Y$). Further, assume that $Q_{XY}$ is cycle-free.

Under the above conditions, if $\mathbf{x} \in T^{(n)}(Q_X)$, $\mathbf{y} \in T^{(n)}(Q_Y)$, and $Q_{X,Y}(\mathbf{x}, \mathbf{y}) > 0$, then $(\mathbf{x}, \mathbf{y}) \in T^{(n)}(Q_{XY})$.

*Proof:* Let $N(a, \mathbf{x})$ denote the number of occurrences of a symbol $a \in \mathcal{X}$ in the sequence $\mathbf{x}$. Likewise, define $N((a, b), (\mathbf{x}, \mathbf{y}))$ to be the number of simultaneous occurrences of the pair $(a, b)$ in the sequence $(\mathbf{x}, \mathbf{y})$. To prove the lemma, we apply induction on the size of $\mathcal{X} \times \mathcal{Y}$. The smallest non-trivial case corresponds to $|\mathcal{X} \cup \mathcal{Y}| = 3$, i.e., either $|\mathcal{X}| = 2$ and $|\mathcal{Y}| = 1$ or $|\mathcal{X}| = 1$ and $|\mathcal{Y}| = 2$. In the first case, $N((a, b), (\mathbf{x}, \mathbf{y})) = N(a, \mathbf{x})$ for all $(a, b) \in \mathcal{X} \times \mathcal{Y}$. Thus, $\mathbf{x} \in T^{(n)}(Q_X)$ implies that $(\mathbf{x}, \mathbf{y}) \in T^{(n)}(Q_{XY})$. A similar argument holds for the second case.

Assume that the lemma is true whenever $|\mathcal{X} \cup \mathcal{Y}| < K$. Suppose now that for $Q_{XY}$, $|\mathcal{X} \cup \mathcal{Y}| = K$. Notice that if $Q_{XY}$ has no cycles, then the connectivity graph $G(Q_{XY})$ on $\mathcal{X} \cup \mathcal{Y}$ has at least one vertex with exactly one edge connected to it. To see this, pick any vertex $v_1$ in $\mathcal{X} \cup \mathcal{Y}$ and construct a sequence of vertices $v_1, v_2, \ldots$ such that $(v_i, v_{i+1})$ are pairs of connected vertices and $v_i \neq v_{i+2}$ for each $i \geq 1$. Since $\mathcal{X} \cup \mathcal{Y}$ is a finite set, either $v_j = v_1$ for some $j > 1$ or the sequence terminates at a vertex $v_k$ which has exactly one edge connected to it. Since $Q_{XY}$ is cycle-free, it follows that the second condition must be true. Further, the vertex $v_k$ also satisfies the property that the transition probability from $v$ to its neighbour is 1 under $Q_{XY}$.

Fix any $\mathbf{x} \in T^{(n)}(Q_X)$ and $\mathbf{y} \in T^{(n)}(Q_Y)$ such that $Q_{XY}(\mathbf{x}, \mathbf{y}) > 0$. Let $v$ be a vertex in $G(Q_{XY})$ that has exactly one connected edge. The following argument shows that $\mathbf{x}, \mathbf{y} \in T^{(n)}(Q_{XY})$. Since the argument is symmetrical in $X$ and $Y$, without loss of generality, assume that $v \in \mathcal{X}$ and $w \in \mathcal{Y}$ be the vertex connected to $v$ in $G(Q_{XY})$. Since $\mathbf{x} \in T^{(n)}(Q_X)$, $N(v, \mathbf{x}) = nQ_X(v)$ and therefore, $N((v, w), (\mathbf{x}, \mathbf{y})) = nQ_X(v) = nQ_{X,Y}(v, w)$.

Now, let $\mathcal{X}' = \mathcal{X} \setminus \{v\}, \mathcal{Y}' = \mathcal{Y}$, and $Q'_Y = Q_Y$. Define probability mass functions $Q'_X$ on $\mathcal{X}'$ and $Q'_{XY}$ on $\mathcal{X}' \times \mathcal{Y}'$ as

follows:

$$\begin{aligned} Q'_X(x) &\triangleq Q_X(x)/(1-Q_X(v)) \; \forall \; x \in \mathcal{X}' \text{ and} \\ Q'_{XY}(x,y) &\triangleq Q_{XY}(x,y)/(1-Q_{XY}(v,w)) \\ &\qquad \forall \; (x,y) \in \mathcal{X}' \times \mathcal{Y}'. \end{aligned}$$

Let $(\mathbf{x}', \mathbf{y}')$ be the subsequence of $(\mathbf{x},\mathbf{y})$ of length $n' = n - N(v,\mathbf{x})$ obtained by deleting the indices that correspond to occurrences of $(v,w)$ in $(\mathbf{x},\mathbf{y})$. It can be verified that $\mathbf{x}' \in T_X^{(n')}(Q'_X)$ and $\mathbf{y}' \in T_Y^{(n')}(Q'_Y)$. Since, $Q'_{XY}$ is cycle-free and $|\mathcal{X}' \cup \mathcal{Y}'| = K-1$, by the induction hypothesis, $(\mathbf{x}',\mathbf{y}') \in T^{(n')}(Q'_{XY})$. Hence, $\forall (x,y) \in \mathcal{X} \times \mathcal{Y} \setminus \{(v,w)\}$,

$$\begin{aligned} N((x,y),(\mathbf{x},\mathbf{y})) &= N((x,y),(\mathbf{x}',\mathbf{y}')) \\ &= n' Q'_{XY}(x,y) \\ &= (n - nQ_{XY}(v,w))Q_{XY}(x,y) \times \\ &\qquad 1/(1-Q_{XY}(v,w)) \\ &= nQ_{XY}(x,y). \end{aligned}$$

This shows that $(\mathbf{x},\mathbf{y}) \in T^{(n)}(Q_{XY})$. ∎

*Lemma 3:* Let $f: \mathcal{Y} \to \mathcal{Z}$ be such that $H(X|Y) = H(X|f(Y))$. Then, $\mathcal{R}_{\mathcal{Z}}(X,Y) = \mathcal{R}_{\mathcal{Z}}(X,f(Y))$.

*Proof:* Clearly, $\mathcal{R}_{\mathcal{Z}}(X,Y) \supseteq \mathcal{R}_{\mathcal{Z}}(X,f(Y))$ since $\mathcal{E}_y$ can compute $f(Y)$ and hence hence, operate at all rate points in $\mathcal{R}_{\mathcal{Z}}(X,f(Y))$. To see the reverse inclusion, define a $\mathcal{Y}$-valued random variable $Y'$ satisfying the Markov chain $Y - f(Y) - Y'$ and let $p_{Y'|f(Y)}(y|z) = p_{Y|f(Y)}(y|z)$ for all $(y,z) \in \mathcal{Y} \times \mathcal{Z}$. It follows that $H(X|Y) = H(X|Y')$, $p_Y = p_{Y'}$, and therefore, $p_{X|Y}(x|y) = p_{X|Y'}(x|y)$. Hence, the joint distribution of $X$ and $Y$ is same as that of $X$ and $Y'$, which implies that $\mathcal{R}_{\mathcal{Z}}(X,Y) = \mathcal{R}_{\mathcal{Z}}(X,Y')$. Finally, note that given $f(Y)$, $\mathcal{E}_y$ can generate $Y'$ randomly. Therefore, $\mathcal{R}_{\mathcal{Z}}(X,f(Y)) \supseteq \mathcal{R}_{\mathcal{Z}}(X,Y') = \mathcal{R}_{\mathcal{Z}}(X,Y)$. ∎